\documentclass[dvipdfm,conference]{IEEEtran}
\IEEEoverridecommandlockouts

\usepackage{cite}
\usepackage{amsmath,amssymb,amsfonts}
\usepackage{algorithmic}

\usepackage{textcomp}
\def\BibTeX{{\rm B\kern-.05em{\sc i\kern-.025em b}\kern-.08em
    T\kern-.1667em\lower.7ex\hbox{E}\kern-.125emX}}
\usepackage{tabularx}
\usepackage[pdftex]{graphicx}

\newcommand{\balpha}{\boldsymbol{\alpha}}

\newcommand{\bmu}{\boldsymbol{\mu}}

\newcommand{\bA}{\boldsymbol{A}}

\newcommand{\bH}{\boldsymbol{H}}
\newcommand{\bI}{\boldsymbol{I}}

\newcommand{\bR}{\boldsymbol{R}}

\newcommand{\bU}{\boldsymbol{U}}

\newcommand{\bW}{\boldsymbol{W}}

\newcommand{\ba}{\boldsymbol{a}}

\newcommand{\bh}{\boldsymbol{h}}

\newcommand{\bw}{\boldsymbol{w}}

\newcommand{\bzero}{\boldsymbol{0}}

\newcommand{\bVb}{{\tilde \bVb}}

\newcommand{\cS}{{\mathcal S}}

\newcommand{\cCN}{\mathcal{CN}}

\newcommand{\T}{\tiny \mbox{T}}  
\newcommand{\R}{\tiny \mbox{R}}

\newcommand{\aod}{{\bf \phi}^{\T}}

\newcommand{\aoa}{{\bf \phi}^{\R}}
\newcommand{\TF}{\tiny \mbox{TF}}

\newcommand{\DR}{\mbox{DR}}

\newcommand{\aodn}{{\bf \theta}^{\T}}

\newcommand{\aoan}{{\bf \theta}^{\R}}

\begin{document}

\title{A Framework for Developing Algorithms for Estimating Propagation Parameters from Measurements
\thanks{Sayeed's work was partly supported by the US NSF through grants \#1703389 and \#1629713.}}

\author{\IEEEauthorblockN{\textsuperscript{1}Akbar Sayeed, \textsuperscript{2}Peter Vouras, \textsuperscript{2}Camillo Gentile, \textsuperscript{2}Alec Weiss, \textsuperscript{2}Jeanne Quimby, 
\textsuperscript{3}Zihang Cheng, \textsuperscript{3}Bassel Modad, \\ \textsuperscript{3}Yuning Zhang, \textsuperscript{4}Chethan Anjinappa, \textsuperscript{4}Fatih Erden,  \textsuperscript{4}Ozgur Ozdemir, \textsuperscript{5}Robert M\"{u}ller, \textsuperscript{5}Diego Dupleich, \textsuperscript{5}Han Niu, \\
\textsuperscript{6}David Michelson, \textsuperscript{6}Aidan Hughes}
\and
\IEEEauthorblockA{\textsuperscript{1}\textit{Electrical  \& Computer Engineering}\\
\textit{U. Wisconsin}\\
Madison, WI\\
akbar.sayeed@wisc.edu} 
\and
\IEEEauthorblockA{\textsuperscript{2}\textit{Communications Technology Laboratory}\\
\textit{National Institute of Standards \& Technology}\\
Gaithersburg, MD and Boulder, CO\\
\{peter.vouras, camillo.gentile\}@nist.gov \\
\{alec.weiss, jeanne.quimby\}@nist.gov}
\and
\IEEEauthorblockA{\textsuperscript{3}\textit{Electrical \& Computer Engineering}\\
\textit{U. Southern California}\\
Los Angeles, CA\\
\{zihangch, aboualim, yzhang26\}@usc.edu}
\and
\IEEEauthorblockA{\textsuperscript{4}\textit{Electrical \& Computer Engineering}\\
\textit{NC State U.}\\
Raleigh, NC\\
\{canjina, ferden, oozdemi\}@ncsu.edu}
\and
\IEEEauthorblockA{\textsuperscript{5}\textit{Technical U.  Ilmenau}\\
Ilmenau, Germany\\
mueller.robert@tu-ilmenau.de, \\
\{diego.dupleich, niu.han\}@tu-ilmenau.de}
\and
\IEEEauthorblockA{\textsuperscript{6}\textit{Electrical \& Computer Engineering}\\
\textit{U. British Columbia}, Vancover, CA\\
davem@ece.ubc.ca, \\
 aidan.hughes@alumni.ubc.ca}
}

\maketitle

\begin{abstract}
A framework is proposed for developing and evaluating algorithms for extracting multipath propagation components (MPCs) from measurements collected by sounders at millimeter-wave (mmW) frequencies. To focus on algorithmic performance, an idealized model is proposed for the spatial frequency response of the propagation environment measured by a sounder. The input to the sounder model is a pre-determined set of MPC parameters that serve as the ``ground truth''. A three-dimensional angle-delay (beamspace) representation  of the measured spatial frequency response serves as a natural domain for implementing and analyzing MPC extraction algorithms. Metrics for quantifying the error in estimated MPC parameters are introduced. Initial results are presented for a greedy matching pursuit algorithm that performs a least-squares (LS) reconstruction of the MPC path gains within the iterations. The results indicate that the simple greedy-LS algorithm has the ability to extract MPCs over a large dynamic range,  and suggest several avenues for further performance improvement through extensions of the greedy-LS algorithm as well as by incorporating features of  other algorithms, such as SAGE and RIMAX. 
\end{abstract}

\section{Introduction}
\label{sec:intro}
Accurate modeling of the multipath propagation environment is critical for the design and deployment of wireless networks, especially at mmW and Terahertz (THz) frequencies that are part of 5G and emerging standards. Accurate channel modeling in turn relies on appropriate and accurate measurements of the propagation environment collected by sounders. Wideband directional sounders at mmW and THz frequencies can take on various forms depending  on the beamforming mechanism, such as phased arrays, lens arrays or mechanically pointed horn antennas  \cite{ch_ams:brady_taps:12, ch_ams:sayeed_gcom:16}, which in turn dictates different approaches for measuring and calibrating sounder characteristics. The final performance of the MPC extraction thus depends on both the sounder hardware characteristics as well as the estimation algorithms used for processing the sounder measurements. The focus of this paper is on the development and evaluation of MPC estimation algorithms while assuming an ``idealized'' model for the sounder measurements that ignores typical hardware non-idealities, such as imperfections in array patterns and frequency response.  This work is part of the ongoing work by the NIST 5G Channel Model Alliance and builds on a recent work on benchmarking different sounders being utilized by the Alliance members \cite{nist_twc:20}. 
\begin{figure}[htb]
\centerline{\includegraphics[width=3.6in]{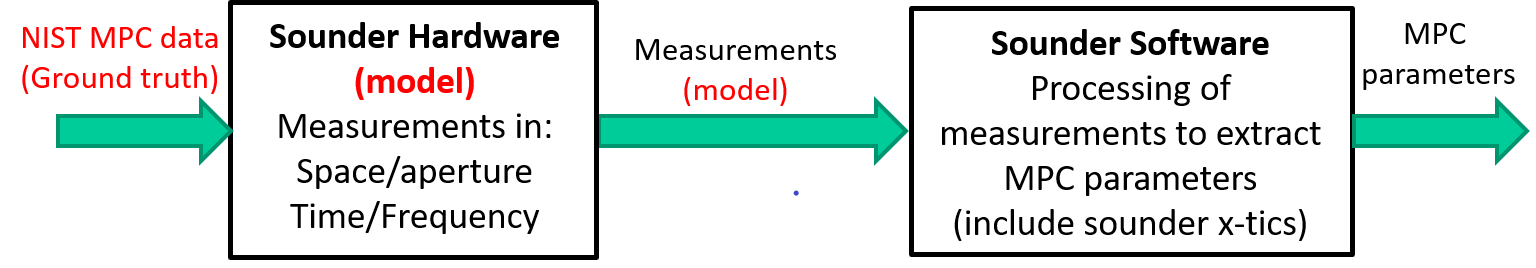}} 
\caption{\footnotesize{\sl Illustration of the proposed methodology for evaluation of MPC parameter extraction algorithms from ``idealized'' channel measurements.  An  ``idealized'' mathematical model for sounder is used to synthetically (computationally) generate the sounder output - the spatial frequency response matrix of the propagation environment - for a given set of known MPC parameters, which serve as  the ``ground truth''.  The idealized synthetic spatial frequency response of the channel serves as the input to the MPC parameter estimation algorithm. }}
\label{fig:method}
\end{figure}
The proposed framework for developing  MPC extraction algorithms from sounder measurements is illustrated in Fig.~\ref{fig:method}.  The measurements collected by the sounder are processed to estimate the MPCs of the propagation channel. The estimation algorithms in general also incorporate the critical characteristics of the sounder that impact the measurements, including the frequency response, array patterns, etc. \cite{nist_twc:20}. In the proposed ``idealized'' evaluation of MPC estimation algorithms, the actual channel measurements are replaced with an ``idealized'' mathematical model for the sounder as illustrated in Fig.~\ref{fig:method}.  A known set of MPC parameters is provided as an input to the sounder model which then generates the corresponding spatial frequency response matrix as the output. The MPC parameter estimation algorithms are then applied to the spatial frequency response matrix generated by the sounder model.  In this work, the ``ground truth'' MPC parameters used as an input to the sounder model are those provided by NIST using their sophisticated sounders and algorithms \cite{nist_mpc:19}.

\section{Physical Channel Model and Its ``Beamspace'' Sampled Representation}
\label{sec:chan_mod}
This section describes a physical model for the multipath propagation channel and its sampled beamspace representation, induced by the key physical parameters of the sounder, to develop an ``idealized'' model for the sounder measurements. 
\subsection{Physical Channel Model for Uniform Linear Arrays}
\label{sec:phy_ams}
Consider a  sounder in which the transmitter (TX) and the receiver (RX) are equipped with uniform linear arrays (ULAs).  In the static scenario, the physical model can be expressed as a spatial frequency response  matrix \cite{ch_ams:bonek:01,ch_ams:cost2100:12,ch_ams:sayeed:02,ch_ams:sayeed_book:08}
\begin{align}
\bH(f) = \sum_{n=1}^{N_p} \alpha_n \ba^{\R}(\aoan_{n})  \ba^{\T \dagger}(\aodn_{n}) e^{-j2\pi \tau_n f} 
\label{Hf} 
\end{align}
which represents a MIMO (multiple input multiple output) channel connecting a TX ULA with $N^{\T}$ antennas and a RX ULA with $N^{\R}$ antennas. The channel is represented by the  $N^{\R} \times N^{\T}$  spatial frequency response matrix $\bH(f)$  which captures the signal propagation over $N_p$ paths, with $\alpha_n$,  $\aodn_n$,  $\aoan_n$, and $\tau_n$  denoting the complex amplitude, angle of departure (AoD), angle of arrival (AoA), and delay of the $n$-th path.  The AoA’s and AoD’s  in (\ref{Hf}) represent the spatial frequencies that are induced by the physical angles, $\aoa_{n}$ and  $\aod_{n}$, defined with respect to the broadside direction, via the relationship: 
\[
\theta=\frac{{d}}{\lambda}  \sin(\phi) =  \frac{1}{2}  \sin(\phi)
\]
where $d$ is the antenna spacing, $\lambda$ is the operating wavelength, and the second equality corresponds to half-wavelength (critical) antenna spacing. The vector  
$\ba^{\T}(\aodn)$ is an $N^{\T} \times 1$ steering vector at the TX for sending a signal in the direction 
$\aodn$, and $\ba^{\R}(\aoan)$  is an $N^{\R} \times 1$ response vector of the RX array for a signal arriving from the direction $\aoan$. For ULAs, the array steering and response vectors take the form of discrete spatial sinusoids with frequencies $\aodn, \aoan \in [-0.5,0.5]$  \cite{ch_ams:bonek:01,ch_ams:sayeed:02,ch_ams:sayeed_book:08}:
\begin{eqnarray}
\ba^{\T}(\aodn) &=& \left [ 1, e^{j2\pi \aodn}, \cdots, e^{j2\pi \aodn(N^{\T} - 1)} \right ]^{*\dagger} \\ 
\ba^{\R}(\aoan) &=& \left [ 1, e^{j2\pi \aoan}, \cdots, e^{j2\pi \aoan (N^{\R} - 1)} \right ]^{*\dagger}
\label{st_vec}  
\end{eqnarray}
where the superscript $^\dagger$ denotes the Hermitian (complex conjugate) transpose and $^*$ denotes complex conjugation.

The model (\ref{Hf}) is widely used for simulating wireless channels. However, it assumes knowledge of the MPC parameters at perfect (infinite) angle-delay resolution. On the other hand, any sounder/system in practice has a finite resolution in angle-delay, which also impacts the statistical characteristics of the estimated MPC parameters  \cite{ch_ams:sayeed_book:08,ch_ams:brady_taps:12}. These challenges are accentuated at mmW frequencies due to: i) the lack of sufficient measurements in different operational environments, and ii) limited capabilities of existing channel sounders, e.g, low spatial resolution and/or mechanical beam pointing. Fundamentally, many technical issues need to be addressed for estimating the angle-delay  MPC parameters from sounder measurements, especially for sounders with antenna arrays for directional measurements. The ``beamspace'' sampled channel representation in angle-delay discussed next provides a useful tool for developing and comparing MPC extraction algorithms.

\subsection{Beamspace Sampled Representation of the Physical Model}
\label{sec:sampled_ams}
A fundamental connection between the measurements made in practice and the physical model above (with continuous parameters) is revealed by a sampled representation of the idealized model (\ref{Hf}) induced by three key parameters of the channel sounder: i) temporal (delay) resolution,  ii) spatial resolution at the TX, and iii) spatial resolution at the RX
\begin{equation}
\Delta \tau = \frac{1}{W} \ ; \ \Delta \aodn =  \frac{1}{N^{\T}} \ ; \ \Delta \aoan = \frac{1}{N^{\R}}
\label{samp_res}
\end{equation}
where $W$ is the (two-sided) bandwidth and the spatial resolutions are for critically spaced antennas \cite{ch_ams:sayeed_book:08}.   The sampled representation of the physical model (\ref{Hf}) is given by
\begin{equation}
\bH(f) \! = \! \sum_{i=1}^{N^{\R}}\sum_{k=1}^{N^{\T}}\sum_{\ell=0}^{L}   H_v(i,k,\ell) 
\ba^{\R}(i \Delta \aoan)\ba^{\T \dagger}(k \Delta \aodn) e^{-j2\pi \ell \Delta \tau f}  
\label{Htf_samp}
\end{equation}
where $L$  represents the maximum number of resolvable delays within the delay spread  $\tau_{max}$: $L  = \left \lceil  \frac{\tau_{max}}{\Delta \tau} \right \rceil = \left \lceil  \tau_{max} W \right \rceil$; 
$\lceil \cdot \rceil$ is the ``ceiling'' operation. The sampled representation is  characterized by the angle-delay (virtual) channel coefficients, $\{ H_v(i,k,\ell) \}$, which can be computed from $\bH(f)$ as \cite{ch_ams:sayeed:02,ch_ams:sayeed_book:08}  
\begin{equation}
H_v(i,k,\ell)  \!  = \! \frac{1}{\scriptstyle{WN^{\R} N^{\T}}}   \int_{\scriptstyle{-\frac{W}{2}}}^{\scriptstyle{\frac{W}{2}}}  \hspace{-5mm} \ba^{\R \dagger}(i \Delta \aoan) \bH(f) \ba^{\T}(k \Delta \aodn) 
  e^{j 2\pi \ell \Delta \tau f} df 
\label{Hv_del_dopp}
\end{equation} 
In essence, the sampled representation (\ref{Htf_samp}) is a three-dimensional (3D) Fourier series expansion of the spatial frequency response matrix $\bH(f)$ in terms of spatial and spectral sinusoids, with the sampled (angle-delay) channel coefficients  in (\ref{Hv_del_dopp}) serving as the expansion coefficients.  The sampled representation is an equivalent representation  
of $\bH(f)$ over the bandwidth $W$ and contains all information about it.

The spatial transformation and sampling in (\ref{Htf_samp}) and (\ref{Hv_del_dopp}) induces an equivalent {\em beamspace representation} of $\bH(f)$ 
 \begin{equation}
\bH_b(f) = \bU^{\R \dagger} \bH(f) \bU^{\T} \Longleftrightarrow \bH(f)= \bU^{\R} \bH_b(f) \bU^{\T \dagger}   
\label{H_b}
\end{equation}
where $\bH_b(f)$ is the beamspace representation, and the matrices $\bU^{\R}$ and $\bU^{\T}$ represent the spatial Discrete Fourier Transform (DFT) matrices, whose columns are steering/response vectors (\ref{st_vec}) for uniformly spaced directions, that map the antenna domain into the angle domain (beamspace):
$\bU^{\T} = \frac{1}{\sqrt{N^{\T}}} \left [ \ba^{\T}(\Delta \theta^{\T}), \ba^{\T}(2\Delta \theta^{\T}), \cdots, \ba^{\T}(N^{\T}\Delta \theta^{\T}) \right ]$ and similarly for $\bU^{\R}$.
The beamspace channel representation in angle-delay is particularly useful at mmW frequencies due to the highly directional nature of propagation. It is a natural domain for representing channel measurements made with directional antennas; e.g., phased arrays, lens arrays or horn antennas \cite{ch_ams:sayeed_gcom:16}.

\section{Estimation of MPCs from Measurements}
\label{sec:extraction}
The focus of this paper is on the idealized model (\ref{Hf}) for a static, frequency-selective MIMO channel. The ``ground truth" MPC data consists of the physical MPC parameters
\begin{equation}
\{\alpha_n, \tau_n, \aoan_n, \aodn_n \ : \ n=1, \cdots, N_p\} 
\label{mpc}
\end{equation}
which are plugged into (\ref{Hf}) to computationally generate synthetic measurements of $\bH(f)$, which are denoted by $\bH_{ms}(f)$.  In practice, the measurements are corrupted by noise
\begin{equation}
\bH_{ms}(f) = \bH(f) + \bW(f)
\label{noisy_meas}
\end{equation}
where $\bW(f)$ denotes the noise matrix, which is assumed to be Additive White Gaussian Noise (AWGN); the entries of the matrix are statistically independent across the different antenna pairs and each entry $W_{i,k}(f)$ represents an AWGN process with unit power spectral density.  The estimation algorithms process $\bH_{ms}(f)$ to generate an estimate of the MPC parameters
\begin{equation}
\{{\hat \alpha}_n, {\hat \tau}_n, {\hat \theta}^{\R}_n, {\hat \theta}^{\T}_n \ : \ n=1, \cdots, {\hat N}_p\}  \  .
\label{mpc_est}
\end{equation}
It is assumed that direct measurements of $\bH_{ms}(f)$ are available in the frequency domain, given the sounder parameters $N^{\T}$, $N^{\R}$, and $W$.  Specifically, the sounder makes temporal measurements at the Nyquist rate over  the duration $T$,  resulting in a total of $N^{\TF}=\frac{T}{\Delta \tau} = {TW}$  samples. A total of $N^{\R} N^{\T}$ temporal measurements are available for all pairwise combinations of TX and RX antennas to populate  $\bH_{ms}(f)$.  
\vspace{-1mm}
\subsection{Maximum Likelihood Estimation}
\label{sec:ml}
Assuming prior knowledge of $N_p$, the optimal estimate of the MPC parameters is the maximum likelihood (ML) estimate
\begin{align}
& \{{\hat \alpha}_n, {\hat \tau}_n,{\hat \theta}^{\R}_n,{\hat \theta}^{\T}_n\} =  \label{mpc_ml}   \\
& \arg \min_{\alpha_n,\tau_n,\aoan_n,\aodn_n}  
 \left \|\bH_{ms}(f) -  \sum_{n=1}^{N_p} \alpha_n \ba^{\R}(\aoan_n)  \ba^{\T \dagger}(\aodn_{n})e^{-j2\pi \tau_n f} \right \|^2  \nonumber
\end{align}
which operates in a high-dimensional spatio-temporal signal space of dimension $D=N^{\T} N^{\R} T W$  and is computationally expensive. The ML estimation in (\ref{mpc_ml}) essentially corresponds to a brute-force search over the $4N_p$ continuous-valued parameters $\{ \alpha_n,\aoan_n, \aodn_n, \tau_n: n=1, \cdots, N_p\}$ that involves $D$-dimensional vectors. For example, for a sounder equipped  with  ULAs at the TX and RX of dimension $N^{\T}=N^{\R} = 35$,  corresponding to 7.4'' critically spaced arrays at 28 GHz, a bandwidth of $W=1$ GHz, and measurement duration $T = 128$ ns, the dimension is $D= 156800$. Proposed algorithms, such as CLEAN \cite{clean:74}, SAGE \cite{sage:94} and RIMAX \cite{ch_ams:richter:05}, are aimed at taming the computational complexity of ML.

\vspace{-1mm}
\subsection{Greedy Matching Pursuit Estimation}
\label{sec:greedy}
Mapping $\bH_{ms}(f)$ into the 3D AoA-AoD-delay (``beamspace'') domain, as in the sampled representation, is a natural first step in developing MPC  estimation algorithms:
\begin{align}
H_{b,ms}(\theta^{\R}, \theta^{\T},\tau)  = & \frac{1}{\scriptstyle{WN^{\R} N^{\T}}}\int_{-\frac{W}{2}}^{\frac{W}{2}}  \hspace{-3mm} \ba^{\R \dagger}(\theta^{\R}) \bH_{ms}(f) \ba^{\T}(\theta^{\T}) e^{j2\pi f \tau} df  \nonumber \\
= & \sum_{n=1}^{N_p} \alpha_n f_{N^{\R}}(\theta^{\R}-\theta^{\R}_n)  f_{N^{\T}}(\theta^{\T}-\theta^{\T}_{n}) \nonumber \\
&    {\rm sinc}(W(\tau-\tau_n))  \  ,
\label{Hb3d} 
\end{align}
where $H_{b,ms}(\theta^{\R}, \theta^{\T},\tau)$ represents the 3D channel impulse response in AoA-AoD-delay space, ${\rm sinc}(x) = \frac{\sin(\pi x)}{\pi x}$ and $f_N(\theta)$ denotes the Dirichlet sinc function $f_N(x) = \frac{\sin(\pi N x)}{\pi x}$.  A simple and often sufficient sub-optimal approach to MPC parameter estimation is the so-called ``greedy'' matching pursuit approach \cite{mp_mallat:93} which sequentially estimates the dominant MPC components. The matching pursuit algorithm operates on $H_{b,ms}(\theta^{\R}, \theta^{\T},\tau)$ to estimate  $K_{dom} $  MPC components:
\vspace{3mm}

\noindent \underline{ALG  Greedy:}   
\begin{align}
\mbox{FOR} \ k=1:K_{dom} & \nonumber  \\
\{{\hat \theta}^{\R}_k,{\hat \theta}^{\T}_k, {\hat \tau}_k\}  = & 
\arg \max_{\theta^{\R},\theta^{\T},\tau}  \left | H_{b,ms}(\theta^{\R},\theta^{\T},\tau) \right | \nonumber \\
{\hat \alpha}_k   = & H_{b,ms}\left ({\hat \theta}^{\R}_k,{\hat \theta}^{\T}_k, {\hat \tau}_k\right )  \nonumber  \\
H_{b,ms}(\theta^{\R},\theta^{\T},\tau)  \longleftarrow & H_{b,ms}(\theta^{\R},\theta^{\T},\tau)  - {\hat \alpha}_k  f_{N^{\R}}\left (\theta^{\R} - {\hat \theta}^{\R}_k \right) \nonumber \\
& f_{N^{\T}}\left (\theta^{\T} - {\hat \theta}^{\T}_k \right ) {\rm sinc}\left (W (\tau - {\hat \tau}_k \right ) ) \nonumber \\
\mbox{END} \hspace{21mm} &   \label{alg:greedy} 
\end{align}
The reconstructed estimate for $\bH(f)$ and the corresponding mean-square error are given by
\begin{align}
{\hat {\bH}}(f)  =  &  \sum_{k=1}^{K_{dom}}  {\hat \alpha}_k \ba^{\R} \left ( {\hat \theta}^{\R}_k \right ) \ba^{\T \dagger}\left ( {\hat \theta}^{\T}_k \right )  e^{-j 2\pi {\hat \tau}_k f}  \label{H_rec} \\
\epsilon & = \frac{1}{W} \int_{-\frac{W}{2}}^{\frac{W}{2}}  \left \| {\hat {\bH}}(f)  - \bH(f) \right \|^2 df \ .
\label{error}
\end{align}
An over-sampled representation of $H_{b,ms}(\theta^{\R}, \theta^{\T}, \tau)$ is computed via (\ref{Hb3d}) to estimate the dominant MPCs in the greedy algorithm (\ref{alg:greedy}). 

\subsection{Least Squares Reconstruction of MPC Amplitudes}
\label{sec:ls_recon}
Once the AoAs, AoDs and delays for $K_{dom}$ dominant MPCs have been estimated, e.g., using the greedy algorithm, a  least squares (LS) update of the MPC complex amplitudes can be obtained to further refine their values.
The measured space-frequency  response is related to the MPC amplitudes  as
\begin{equation}
\bh_{ms} = \bA_{dom} \balpha
\label{ls_mod}
\end{equation}
where $\bh_{ms}$ is the $N_o = N^{\R} N^{\T} N^{\TF}$ dimensional vector representation (with frequency sampling at $N^{\TF}$ points) of the measurement matrix $\bH_{ms}(f)$,  $\bA_{dom}$ is the $N_o \times K_{dom}$ matrix whose columns are space-frequency basis vectors corresponding to the dominant estimated MPCs:
\begin{align}
\bA_{dom} & = [ \ba_1, \ba_2, \cdots, \ba_{K_{dom}} ] \nonumber \\   
\ba_k  &  = \ba^{\TF}({\hat \tau}_k)  \otimes \left [  \ba^{\T*}({\hat \theta}^{\T}_k) \otimes \ba^{\R}({\hat \theta}^{\R}_k) \right ] 
\label{Adom}
\end{align}
and  $\otimes$ denotes the kronecker product. 
The LS estimate for the vector of complex path amplitudes, $\balpha$,  in (\ref{ls_mod}) is given by 
\begin{align}
{\hat \balpha}_{ls} & = \left ( \bA_{dom}^\dagger \bA_{dom} \right )^{-1} \bA_{dom}^\dagger \bh_{ms}  
\label{ls_est}
\end{align}
Note that $\bA_{dom}^\dagger \bA_{dom}$ is a $K_{dom} \times K_{dom}$ matrix  and is generally invertible as long as the estimated basis vectors are sufficiently distinct in (\ref{Adom}) and $K_{dom} < N_o$ which is guaranteed due to multipath sparsity, especially in high-dimensional channels. A LS update improves the estimate of $\balpha$ since the columns of $\bA_{dom}$ are not orthogonal in general.

\subsection{Path Association}
\label{sec:path}
This section presents a procedure for path association (PA) between the estimated and true physical MPC parameters to assess algorithm performance. The PA procedure  identifies $K_{pa} \leq  \min(N_p,K_{dom}) $ physical and estimated paths which are ``closest'' in the AoA-AoD-delay space.  Let $\cS_{est} = \{ 1, 2, \cdots, K_{dom}\}$ denote the set of indices for the $K_{dom}$ MPCs returned by the greedy-LS algorithm, $\cS_{phy} = \{ 1, 2, \cdots, N_p\}$ denote the set of indices for the physical MPCs, and let $\cS_{pa} = \{ 1, 2, \cdots, K_{pa} \}$ denote the set of indices for the output of the PA procedure.  Let $p:  \cS_{pa} \rightarrow \cS_{phy} $ and $q: \cS_{pa} \rightarrow \cS_{est}$ denote index mappings for the physical and estimated MPCs. The objective is to find the mappings $p$ and $q$ that minimize the cost of PA. For given $p$ and $q$, define the cost metric  as
\begin{align}
C_{tot} (p,q) & = \sum_{k=1}^{K_{pa}} C(p_k,q_k)  P_{mpc}(p_k)  \nonumber \\
C(p_k,q_k)  & =      C_{\theta^{\R}} (p_k, q_k) + C_{\theta^{\T}}(p_k, q_k)  +   C_{\tau}(p_k,q_k)   
\label{cost}
\end{align}
and the individual costs are defined as 
\begin{align}
C_{\theta^{\R}} (p_k,q_\ell) & = \left ( \frac{ \theta^{\R}_{p_k} -{\hat \theta^{\R}}_{q_\ell}}{\Delta \theta^{\R}} \right )^2  \ ; \  C_{\theta^{\T}} (p_k,q_\ell)  = \left ( \frac{ \theta^{\T}_{p_k}- {\hat \theta^{\T}}_{q_\ell}}{\Delta \theta^{\T}} \right )^2  \nonumber \\
C_{\tau}(p_k,q_\ell) & =  \left ( \frac{ \tau_{p_k} -{\hat \tau}_{q_\ell}}{\Delta \tau} \right )^2 
\label{cost_ind}
\end{align}
where  the denominators are the resolutions defined in (\ref{samp_res}),
and $P_{mpc}(p_k)$ is the normalized physical path power $P_{mpc}(p_k) = |\alpha_{p_k}|^2/\left ( \sum_{k=1}^{K_{dom}} |\alpha_k|^2  \right )$.
The normalizations in (\ref{cost_ind}) are introduced to make the size of the individual costs comparable to each other.
The Hungarian algorithm (``matchpairs'' function in MATLAB) is used to find the optimal cost-minimizing mappings. The input to the algorithm are all pair-wise costs, $C(p_k,q_\ell); k = 1, \cdots, N_{p} \ , \ \ell = 1, \cdots, K_{dom}$, 
and an adjustable cost value, $C_{um}$, for unmatched pairs. The algorithm returns the mappings for the $K_{pa} $  associated paths: $p_k$, $q_k$, $k=1, \cdots, K_{pa}$.  The post-PA cost is given by (\ref{cost}).  Prior to doing PA, the physical and estimated MPCs are arranged in descending order of the paths powers and the pre-PA cost is computed using (\ref{cost}) for the first $K_{pa}$  paths.  

 In order to further quantify performance, the MPCs are partitioned into two sets, depending on whether the estimated MPC is within the ``resolution bin'' of the corresponding associated physical MPC. Define the following sets
\begin{align}
\hspace{-4mm} \cS_\tau  & = \{ k: |\tau_{p_k} - {\hat \tau}_{q_k}| \leq \Delta \tau \} ; 
  \cS_{\theta^{\R}}  = \{ k: |\theta_{p_k}^{\R} - {\hat \theta}_{q_k} ^{\R}| \leq \Delta \theta^{\R} \} \nonumber  \\
\hspace{-2mm} \cS_{\theta^{\T} } & =  \{ k: |\theta_{p_k}^{\T} - {\hat \theta}_{q_k} ^{\T}| \leq \Delta \theta^{\T} \} ;   \cS_{joint} = \cS_\tau \cap \cS_{\theta^{\R}} \cap \cS_{\theta^{\T}} 
\label{bin_sets}
\end{align}
$\cS_{joint}$ denotes all paths whose AoAs, AoDs, and delays are jointly within their corresponding resolution bins.

\section{Other MPC Estimation Algorithms}
\label{sec:other}
Let $\psi_n = (\alpha_n, \theta^{\R}_n, \theta^{\T}_n, \tau_k)$, $n=1,\cdots, N_p$, denote the MPC parameters. The greedy part of  the greedy-LS algorithm in (\ref{alg:greedy}) is identical to the CLEAN algorithm.  The MPC estimates from the greedy-LS algorithm can be used for  initializing  the iterations in SAGE and RIMAX. Let ${\hat \psi}^{i}_k$, $k=1, \cdots, K_{dom}$, denote the MPC estimates at the $i$-th iteration and
\begin{align}
{\hat \bH}_{k}^i(f) & = {\hat \alpha}^i_k \ba^{\R}\left ({\hat \theta}^{\R,i}_k\right ) \ba^{\T \dagger}\left ({\hat \theta}^{\T,i}_k \right ) e^{-j2\pi {\hat \tau}^{i}_k f} \label{Hk}
\end{align}  the estimated channel component for the $k$-th MPC at the $i$-th iteration.   A simple path-wise iteration of the SAGE algorithm refines the MPC estimates at the $i$-th iteration via the following two steps ($k=1, \cdots, K_{dom}$):
\begin{align}
\mbox{E-step:} \ & \bH_{res,k}^i(f)  = \bH_{ms}(f)- \sum_{k' \neq k} {\hat \bH}_{k'}^i(f) \label{E_step} \\
\mbox{M-step:} \ &  ({\hat \theta}^{\R,i+1}_k, {\hat \theta}^{\T, i+1}_k, {\hat \tau}^{i+1}_k)  = \arg \max_{\theta^{\R}, \theta^{\T}, \tau} {\hat H}_{b,res,k}^i(\theta^{\R}, \theta^{\T}, \tau) \nonumber \\
 &  {\hat \alpha}^{i+1}_k =  {\hat H}_{b,res,k}^i \left ( {\hat \theta}^{\R,i+1}_k, {\hat \theta}^{\T, i+1}_k, {\hat \tau}^{i+1}_k \right ) \ .\label{M_step}
\end{align}
A stopping criterion for the iterations may be based on the rate of reduction in the reconstruction error in (\ref{error}). 

The RIMAX method offers further refinement over greedy-LS, CLEAN and SAGE algorithms. Specifically, it uses a more general model for the measurements which incorporates diffuse multipath components (DMCs):
\begin{align}
\bh_{ms} & =  \bh_{smc} + \bh_{dmc} + \bw  \label{h_rimax}  \\
 \bh_{smc} & = \bA(\bmu)\balpha ; \  \bh_{dmc} \sim \cCN \left ( \bzero, \bR_{dmc} \right ); \  \bw  \sim  \cCN \left (\bzero,  \bI \right ) \nonumber \\
 \bmu  =  \{ & (\theta^{\R}_k, \theta^{\T}_k, \tau_k): k = 1,\cdots,K_{dom}\} ; \balpha = [\alpha_1,\cdots,\alpha_{K_{dom}}]^{\dagger *}  \nonumber
 \end{align}
 In the above $\bh_{ms}$, $\bh_{smc}$, $\bh_{dmc}$ and $\bw$ are vectorized versions (with sampled frequencies) of the measurement channel matrix $\bH_{ms}(f)$, the channel matrix $\bH_{smc}(f)$ consisting of specular multipath components (SMCs), the DMC channel matrix $\bH_{dmc}(f)$ and the noise matrix $\bW(f)$.   While both $\bH_{smc}(f)$ and $\bH_{dmc}$ take the form (\ref{Hf}), the SMC matrix is modeled as deterministic (with $\bA(\bmu)$ taking the form (\ref{Adom})) and the DMC matrix is modeled statistically through the covariance matrix $\bR_{dmc}$. The general ML estimation problem for (\ref{h_rimax}) is
 \begin{align}
 & ({\hat \bmu}, {\hat \balpha},  {\hat \bR}_{dmc})    = \nonumber \\
 &  \arg \min_{\bmu, \balpha, \bR_{dmc}}  \left  [\bh_{ms} - \bh_{smc} \right ]^{\dagger}  \bR_{dan}^{-1} \left [\bh_{ms}-\bh_{smc} \right ]  +  \ln  |\bR_{dan} |   \nonumber \\
 & \bR_{dan} = \bR_{dmc} + \bI 
 \label{ml_rimax}
 \end{align}
 The RIMAX algorithm starts with an initialization of $\bh_{smc}$, using CLEAN or SAGE, e.g., as well as $\bR_{dmc}$, and then iteratively updates the estimates of both the  specular ($\bmu$ and $\balpha$) and diffuse ($\bR_{dmc}$) components:
 \begin{align}
& \mbox{SMC update}: \mbox{fixed} \  {\hat \bR}_{dmc}   \nonumber \\
 & ({\hat \bmu}, {\hat \balpha}) =   \nonumber \\
 &  \arg \min_{\bmu, \balpha}   [\bh_{ms} - \bh_{smc}(\bmu,\balpha) ]^\dagger {\hat \bR}_{dan}^{-1}  [ \bh_{ms}-\bh_{smc}(\bmu,\balpha) ] \nonumber \\
& \mbox{DMC update}: \mbox{fixed} \  ({\hat \bmu}, {\hat \balpha})   \nonumber\\
 & {\hat \bR}_{dmc} = \nonumber \\
 &   \arg \min_{\bR_{dmc}}   [\bh_{ms}-{\hat \bh}_{smc}  ]^\dagger    \bR_{dan}^{-1}  [\bh_{ms}-{\hat \bh}_{smc} ] + \ln |\bR_{dan}|  \nonumber
 \end{align}
 The greedy-LS/CLEAN algorithms have the lowest complexity, followed by SAGE and then RIMAX, which is most complex. 
 
 \section{Numerical Results}
\label{sec: results}
The MPC extraction framework outlined in this paper is illustrated with a data set provided by NIST \cite{nist_mpc:19}  corresponding  to  propagation in a conference room of dimension 10m x 19m x 3m for 20 different TX and RX locations. The first TX and RX location is considered, and a subset of MPCs  whose AoDs and AoAs in azimuth are within the range between $-90$ and $+90$ degrees (both the TX and RX ULAs facing each other) is utilized. 
The idealized sounder model consists of (horizontal) ULAs at the TX and RX with $N^{\T}  = N^{\R}=35$ elements corresponding to a critically spaced 7.4" ULA at $f_c = 28$ GHz. An operational bandwidth of $W=1$GHz is assumed. 
\begin{figure}[htb]
\vspace{-3mm}
\begin{tabular}{cc}
\hspace{-3mm}
\includegraphics[width=1.7in]{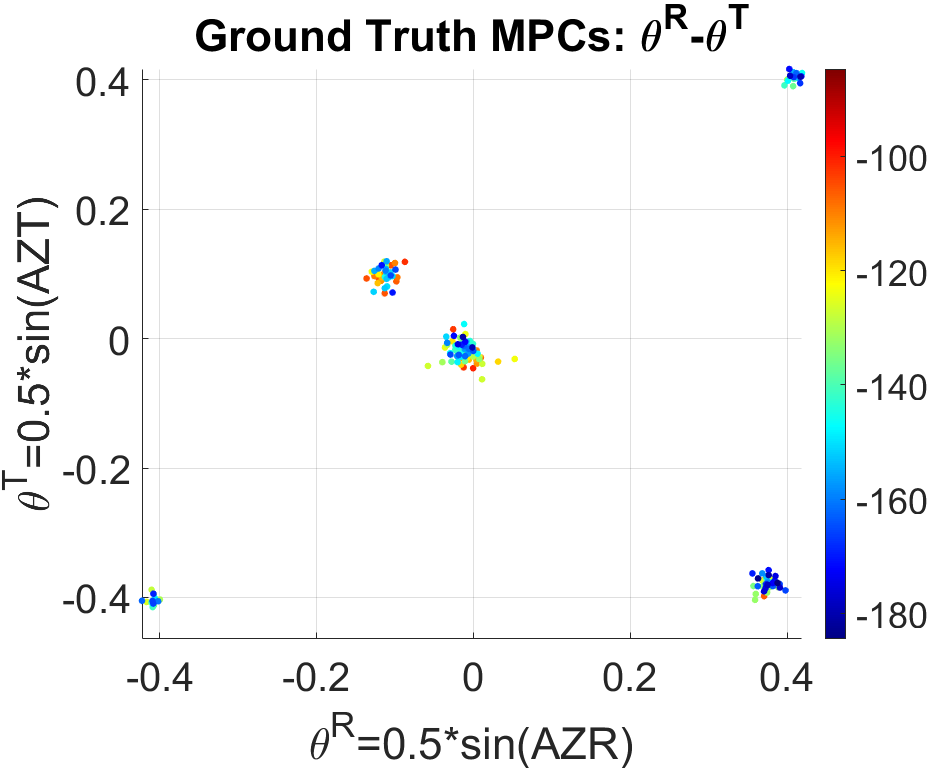}
& 
\hspace{-3mm}
\includegraphics[width=1.7in]{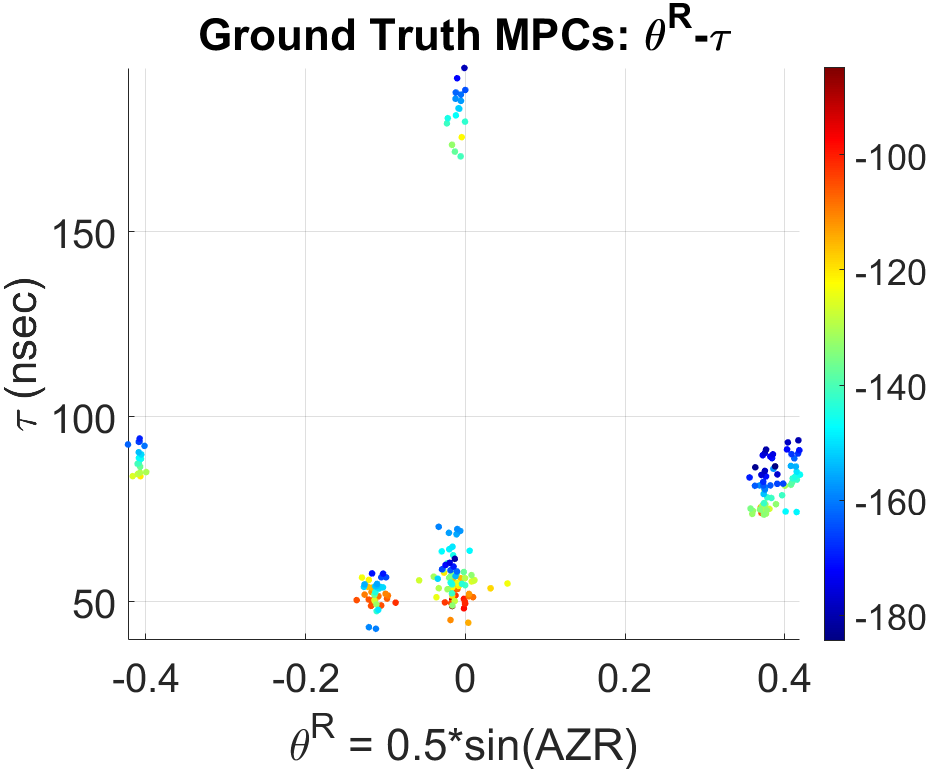}\\
\footnotesize{(a)} & \footnotesize{(b)}
\end{tabular}
\caption{\footnotesize{\sl Plots of $N_p$=$224$  physical (ground truth) MPCs, color coded according to the path gains, retained for $\DR=100$dB. (a) $\theta^{\R}$ versus $\theta^{\T}$. (b) $\theta^R$ versus $\tau$.}}
\label{fig:mpc_phy_est}
\end{figure}

One important parameter is the dynamic range $\mbox{DR}$ of path amplitudes to include in the numerically generated sounder ``measurements". The   $\DR$ is defined as the ratio of the maximum path power to the minimum path power.  Fig.~\ref{fig:mpc_phy_est} plots $\theta^{\R}$ versus $\theta^{\T}$ and $\theta^{\R}$  versus $\tau$  of the $N_p = 224$ (out of 252) ``ground truth'' MPCs that are retained for $\DR=100$dB. The  MPCs are color coded according to their path gain (dB).  Five MPC clusters  are evident from Fig~\ref{fig:mpc_phy_est}.
 \begin{figure}[htb]
 \vspace{-3mm}
\begin{tabular}{cc}
\hspace{-3mm}
\includegraphics[width=1.7in]{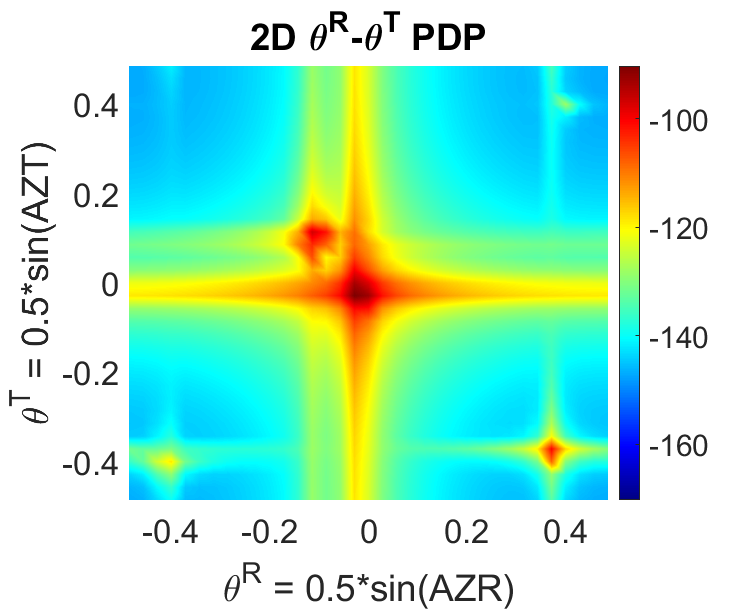} & 
\hspace{-3mm}\includegraphics[width=1.7in]{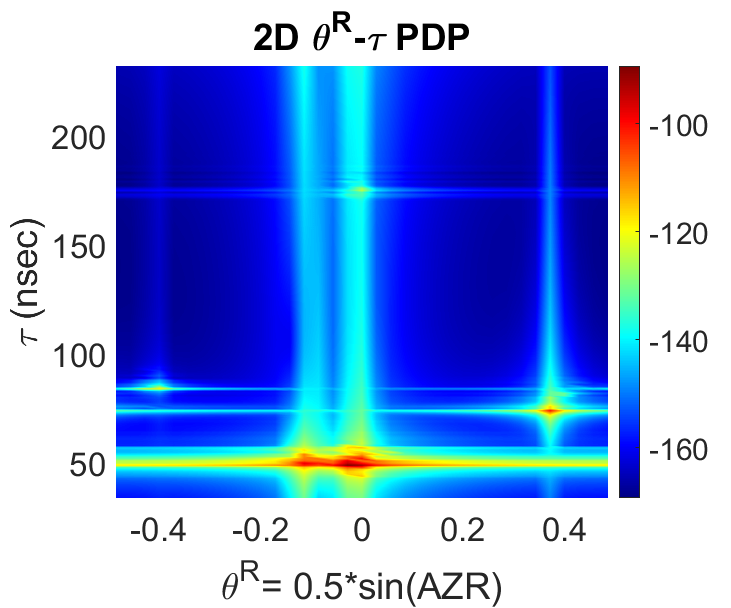} \\
\footnotesize{(a)} & \footnotesize{(b)}
\end{tabular}
\vspace{-2mm}
\caption{\footnotesize{\sl (a) 2D PDP $|H_{b,ms}(\theta^{\R}, \theta^{\T})|^2$. (b)  2D PDP $|H_{b,ms}(\theta^{\R}, \tau)|^2$.  }}
\label{fig:pdp}
\vspace{-3mm}
\end{figure}
The ``measured'' spatial frequency response $\bH_{ms}(f)$ is generated via (\ref{Hf}), using the underlying ``ground truth'' values of the MPC parameters. No measurement noise is considered in this initial benchmarking investigation. 
The 3D beamspace representation $H_{b,ms}(\theta^{\R}, \theta^{\T}, \tau)$  is computed using (\ref{Hb3d}) with $N^{\TF} = 233$ ($T=233$ns). The 2D power distribution profiles (PDPs) in AoA-AoD and AoA-delay are shown in Fig.~\ref{fig:pdp} from which five distinct MPC clusters can be identified. Note that a couple of clusters are associated with weaker MPCs which would become more challenging to estimate in the presence of noise. 

\begin{figure}[htb]
 \vspace{-3mm}
\begin{tabular}{cc}
\hspace{-3mm}
\includegraphics[width=1.7in]{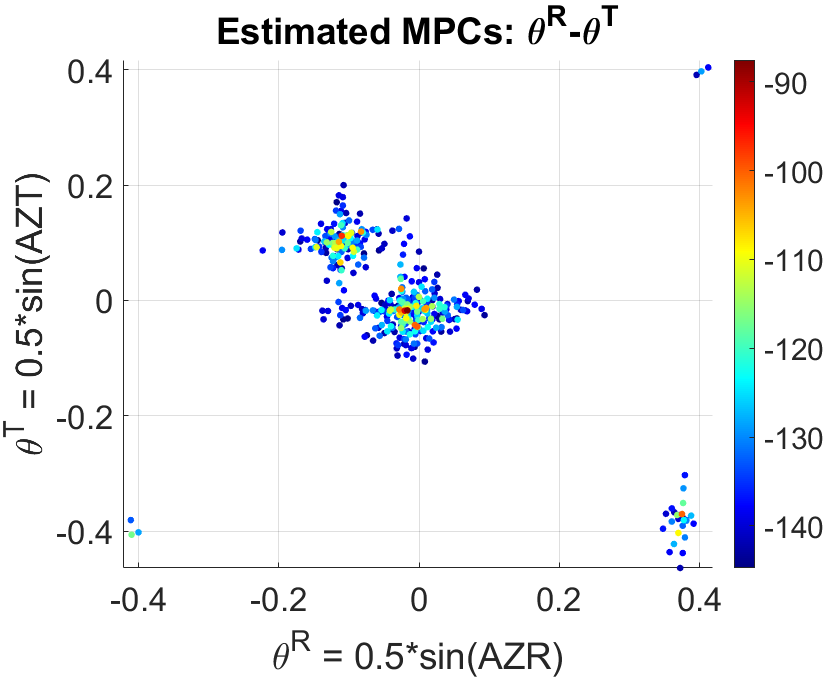} & 
\hspace{-3mm}\includegraphics[width=1.7in]{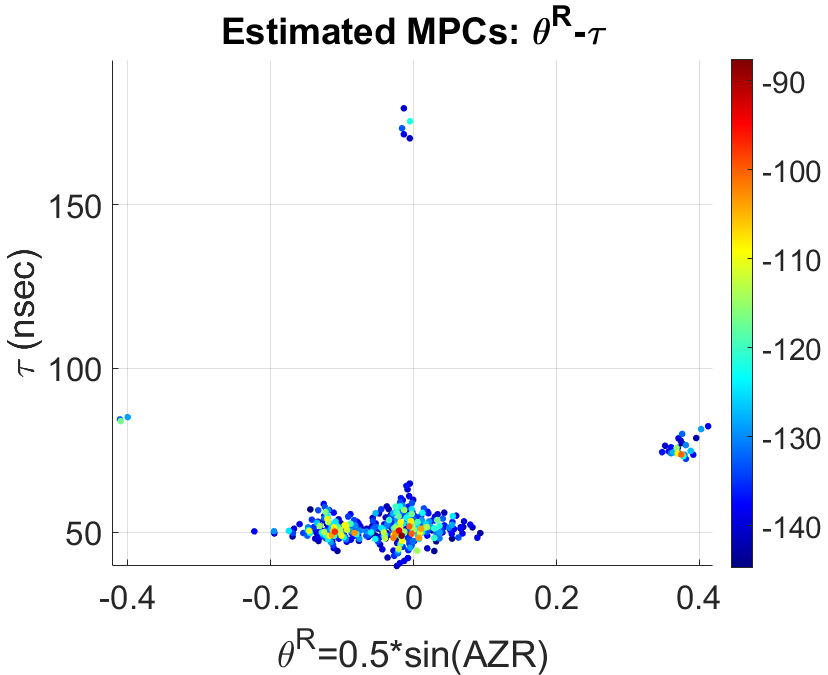} \\
\footnotesize{(a)} & \footnotesize{(b)}
\end{tabular}
\vspace{-2mm}
\caption{\footnotesize{\sl Plots of $K_{dom}$=$448$ estimated MPCs, color coded according to the path gains, returned by the greedy-LS algorithm. (a) $\theta^{\R}$ - $\theta^{\T}$. (b) $\theta^R$ - $\tau$.}}
\label{fig:mpc_est}
\vspace{-1mm}
\end{figure}

The MPC extraction algorithm processes  $H_{b,ms}(\theta^{\R}, \theta^{\T}, \tau)$ to estimate the gains, AoAs, AoDs, and delays of the MPCs. 
We present results for a greedy algorithm that incorporates LS reconstruction within the iterations. Specifically, for each iteration, $K_g=4$ dominant paths are estimated, an LS reconstruction (\ref{ls_est})  is done on all $K_g$ estimated paths, and then the contribution of the $K_{up}=2 (\leq K_g)$  strongest estimated MPCs  is subtracted from $H_{b,ms}(\theta^{\R}, \theta^{\T}, \tau)$ as in (\ref{alg:greedy}) for the update. The algorithm is run for $K_{dom}$=$2N_p$=$448$ iterations and Fig.~\ref{fig:mpc_est} plots $\theta^{\R}$ versus $\theta^{\T}$ and $\theta^{\R}$  versus $\tau$  of the $K_{dom}$ estimated MPCs, again color coded according to the path gains. 
 As evident,  the greedy-LS algorithm does a  good job of estimating the MPCs and identifies paths in each of the five clusters. The normalized reconstruction error (\ref{error}) is 0.034\%. 

\begin{figure}[htb]
 \vspace{-3mm}
\begin{tabular}{cc}
\hspace{-3mm}
\includegraphics[width=1.7in]{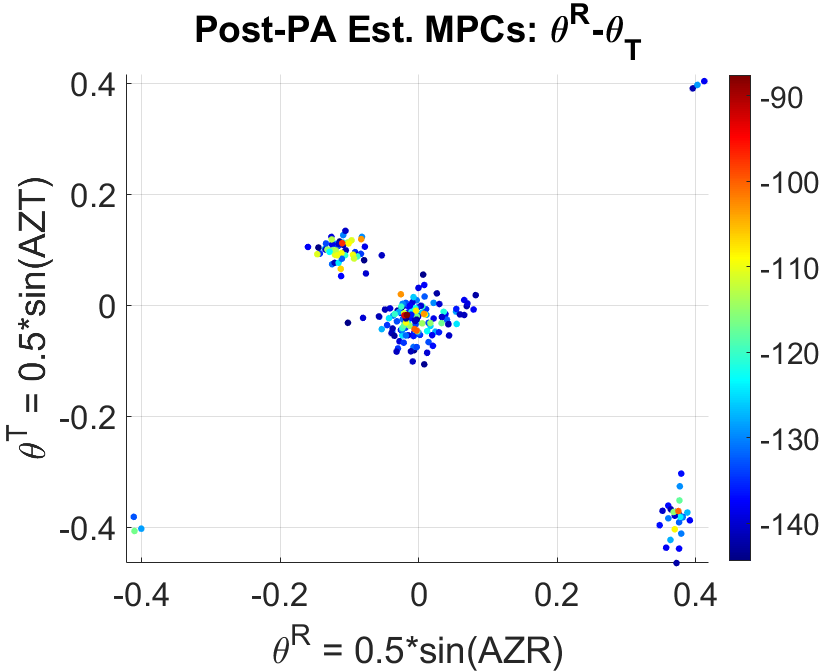} & 
\hspace{-3mm}\includegraphics[width=1.7in]{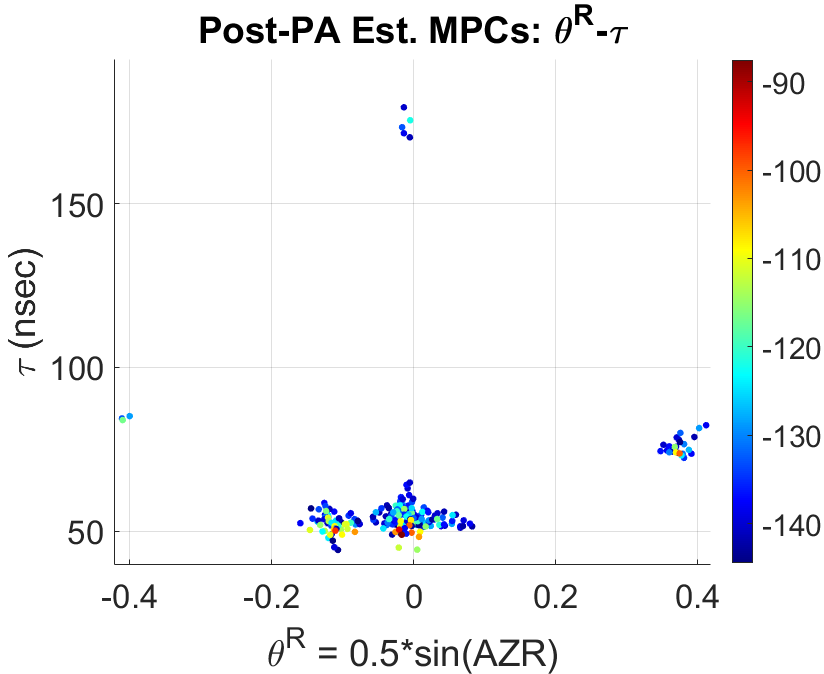} \\
\footnotesize{(a)} & \footnotesize{(b)}
\end{tabular}
\caption{\footnotesize{\sl Plots of $K_{pa}$=$224$ estimated MPCs, color coded according to the path gains, returned by the PA procedure. (a) $\theta^{\R}$ - $\theta^{\T}$. (b) $\theta^R$ - $\tau$.}}
\label{fig:mpc_est_pa}
\vspace{-3mm}
\end{figure}

Since the ground truth values of the MPCs are known {\em a priori}, the estimation errors in AoA, AoD, and delay can be quantified. Path association (PA) is performed between the physical (ground truth) and estimated MPC parameters as discussed in Sec.~\ref{sec:path}. The PA algorithm returns index mappings $p$ and $q$ for  the $K_{pa}=N_p=224$  associated true and estimated MPCs. The post-PA cost metric in (\ref{cost})  is about 0.4\% of the pre-PA cost metric.  Fig.~\ref{fig:mpc_est_pa} plots $\theta^{\R}$ versus $\theta^{\T}$ and $\theta^{\R}$  versus $\tau$  of the $K_{pa}$=$224$ estimated MPCs returned by the PA procedure, again color coded according to the path gains. Compare the difference between Fig.~\ref{fig:mpc_est} and Fig.~\ref{fig:mpc_est_pa} - only the $K_{pa}$ ``best associated'' MPCs from Fig.~\ref{fig:mpc_est} are retained in Fig.~\ref{fig:mpc_est_pa}. 

\begin{figure}[htb]
\begin{tabular}{cc}
\hspace{-3mm}
\includegraphics[width=1.6in]{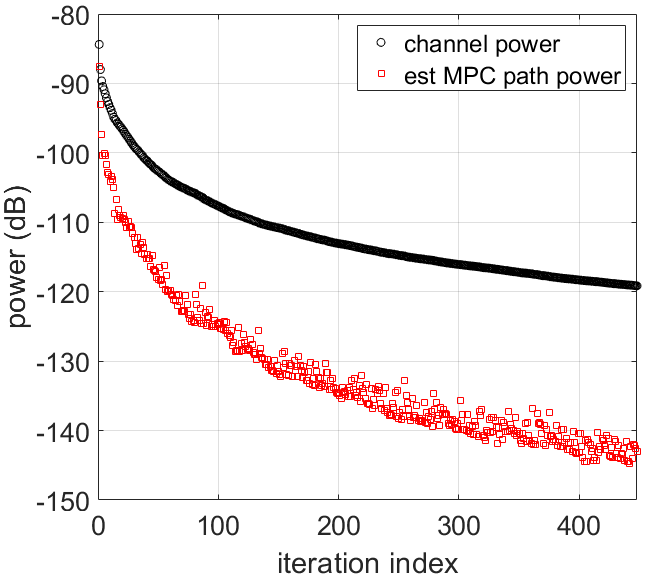} & 
\hspace{-3mm} 
\includegraphics[width=1.7in]{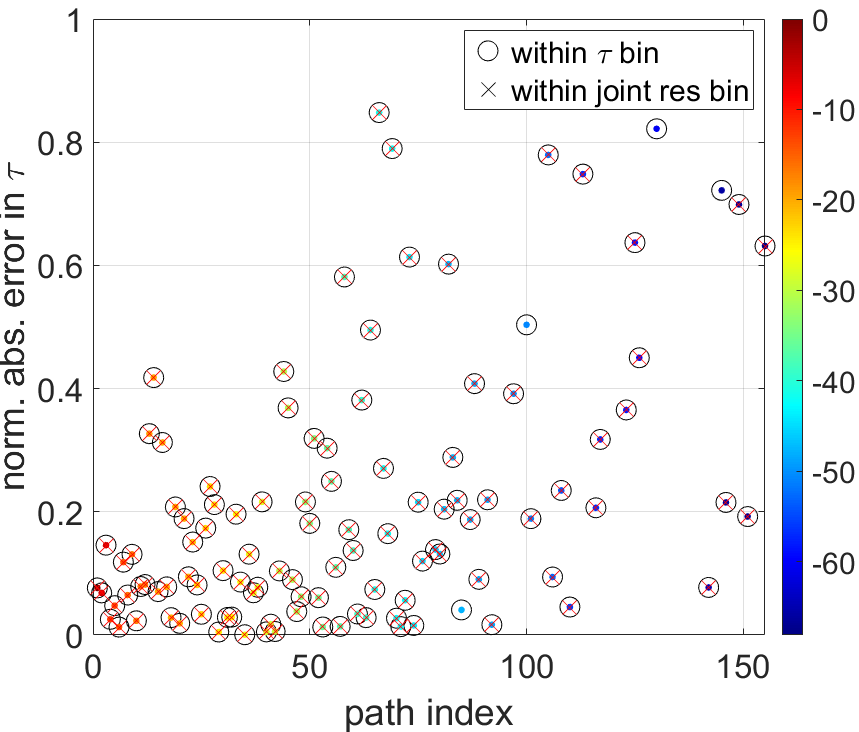}  \\
\footnotesize{(a)} & \footnotesize{(b)} \\
\hspace{-3mm}
\includegraphics[width=1.7in]{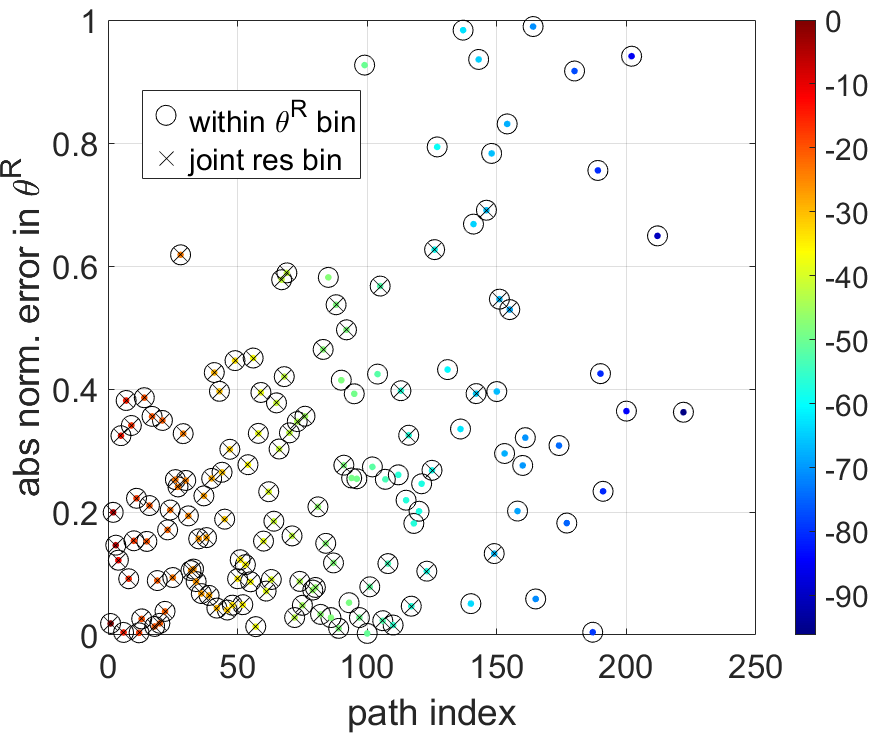}  & 
\hspace{-3mm} 
\includegraphics[width=1.7in]{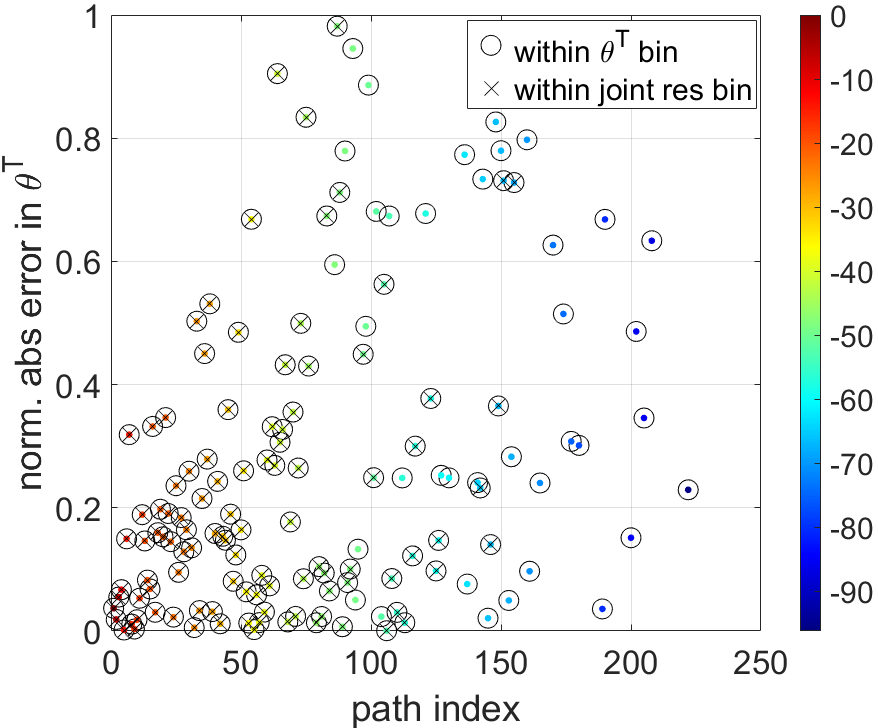} \\
 \footnotesize{(c)} & \footnotesize{(d)} 
\end{tabular}
\vspace{-2mm}
\caption{\footnotesize{\sl (a) Plots of the residual channel power and the power of the dominant estimated path gain as a function of iteration index in the greedy-LS algorithm. (b)-(d) The post-PA normalized absolute error between true  and estimated MPCs for paths that are within resolution bins: (b) $\tau$;  $|\cS_{\tau}| = 108$, (c) $\theta^{\R}$ ;  $|\cS_{\theta^{\R}}| = 148$ , (d) $\theta^{\T}$ ;  $|\cS_{\theta^{\T}}|=142$ . The $|\cS_{joint}|= 104$ MPCs that are jointly within all three resolution bins are identified with a $\times$ marker in (b)-(d).}}
\label{fig:mpc_error_448_224}
\vspace{-3mm}
\end{figure}

Fig.~\ref{fig:mpc_error_448_224}(a)  plots  the channel power and the power of the dominant MPC path gain subtracted as a function of the iteration index. It is important to note that the (residual) channel power is monotonically decreasing with each iteration, as expected.  Figs.~\ref{fig:mpc_error_448_224}(b)-(d)  plot the normalized absolute errors in the estimates of parameters for MPCs which lie within a resolution bin of the associated ground truth MPCs. The estimated MPCs  whose parameters  are jointly within all three resolution bins,  see (\ref{bin_sets}), are also identified with a $\times$ marker.   Note that  $|\cS_{\tau}|$=108, $|\cS_{\theta^{\R}}|$=148,  $|\cS_{\theta^{\T}}|$=142, and  $|\cS_{joint}|$=104. Thus, out of the $N_p=224$ true underlying MPCs, the PA procedure identifies 104 estimated MPCs whose AoAs, AoDs, and delays are jointly within the corresponding resolution bins.  Of course, one cannot rely on PA in practice when there is no information about the ground truth.  But these initial results indicate that the basic greedy-LS algorithm has the ability to extract weak MPCs and could serve as a starting point for applying other methods  such as SAGE and RIMAX.

\section{Conclusion}
\label{sec:conc}
The initial results presented in this paper are promising and suggest many directions for future work. One key question is  how to prune the large number of MPCs extracted to reflect the actual physical paths (without PA). A number of possibilities exist here, including modifications of the greedy-LS algorithm that exploit the concept of path partitioning in the beamspace representation  \cite{ch_ams:sayeed:02,ch_ams:sayeed_book:08}, as well as incorporating features of  more advanced algorithms, such as SAGE \cite{sage:94} and RIMAX \cite{ch_ams:richter:05}.  Another promising direction is to extend the greedy-LS algorithms so that it can be applied in a local region of the angle-delay space, to focus on specific MPC clusters; see Fig. ~\ref{fig:pdp}.  Finally, evaluation of the MPC estimation algorithms in the presence of noise is needed. Addressing the MPC pruning issue will also likely improve the performance in the presence of measurement noise. These issues are currently under investigation by the NIST 5G Alliance in ongoing work.

\bibliographystyle{IEEEtran}
\bibliography{mpc_extraction}

\end{document}